\documentclass[sigconf]{acmart}

\usepackage{amsmath,amsfonts,bm,bbm}
\usepackage{amsthm}
\usepackage{graphicx}
\usepackage{caption}
\usepackage{color}
\usepackage{subfigure}
\usepackage{comment}
\usepackage{enumitem}
\usepackage{mathtools}
\usepackage{thmtools, thm-restate}
\usepackage[ruled,linesnumbered]{algorithm2e}  
\usepackage{algorithmic}
\usepackage{verbatim}

\usepackage{hyperref}
\usepackage{cleveref}
\usepackage{xcolor}
\usepackage{natbib}
\usepackage{listings}
\usepackage{xparse}
\usepackage{blindtext}
\usepackage{lipsum}
\usepackage{caption}
\usepackage{subcaption}
\usepackage{enumitem}
\usepackage{makecell}
\usepackage{anyfontsize}

\theoremstyle{plain} \numberwithin{equation}{section}
\newtheorem{theorem}{Theorem}[section]
\numberwithin{theorem}{section}

\theoremstyle{definition}

\newtheorem{example}[theorem]{Example}

\usepackage[breakable, skins]{tcolorbox}
\definecolor{greyC}{RGB}{180,180,180}
\definecolor{greyL}{RGB}{235,235,235}

\def\1{\mathbbm{1}}

\def\vmu{{\bm{\mu}}}
\def\vtheta{{\bm{\theta}}}

\def\vw{{\mathbf{w}}}

\def\vz{{\bm{z}}}

\def\mX{{\bm{X}}}

\def\mSigma{{\bm{\Sigma}}}

\DeclareMathAlphabet{\mathsfit}{\encodingdefault}{\sfdefault}{m}{sl}
\SetMathAlphabet{\mathsfit}{bold}{\encodingdefault}{\sfdefault}{bx}{n}

\def\gA{{\mathcal{A}}}
\def\gB{{\mathcal{B}}}

\def\gD{{\mathcal{D}}}

\def\gN{{\mathcal{N}}}

\def\gR{{\mathcal{R}}}

\newcommand{\E}{\mathbb{E}}

\newcommand{\R}{\mathbb{R}}

\DeclareMathOperator*{\argmax}{arg\,max}

\newcommand{\hyperzero}{{\sc HyperZero}}

\AtBeginDocument{%
  }

\copyrightyear{2025} 
\acmYear{2025} 
\setcopyright{cc}
\setcctype{by}
\acmConference[KDD '25]{Proceedings of the 31st ACM SIGKDD Conference on Knowledge Discovery and Data Mining V.1}{August 3--7, 2025}{Toronto, ON, Canada}
\acmBooktitle{Proceedings of the 31st ACM SIGKDD Conference on Knowledge Discovery and Data Mining V.1 (KDD '25), August 3--7, 2025, Toronto, ON, Canada}
\acmDOI{10.1145/3690624.3709409}
\acmISBN{979-8-4007-1245-6/25/08}

\begin{document}

\title{\hyperzero: A Customized End-to-End Auto-Tuning System for Recommendation with Hourly Feedback}

\author{Xufeng Cai}
\authornote{Equal contribution.}
\authornote{This work was done when the author was an intern at Meta.}
\email{xcai74@wisc.edu}
\affiliation{%
  \department{Department of Computer Sciences}
  \institution{University of Wisconsin-Madison}
  \city{Madison}
  \state{WI}
  \country{United States}
}

\author{Ziwei Guan}
\authornotemark[1]
\email{guanziwei@meta.com}
\affiliation{%
  \institution{Meta Platforms Inc.}
  \city{Menlo Park}
  \state{CA}
  \country{United States}
}

\author{Lei Yuan}
\authornotemark[1]
\email{leiyuan@meta.com}
\affiliation{%
  \institution{Meta Platforms Inc.}
  \city{Bellevue}
  \state{WA}
  \country{United States}
} 

\author{Ali Selman Aydin}
\authornotemark[1]
\email{aliselmanaydin@meta.com}
\affiliation{%
  \institution{Meta Platforms Inc.}
  \city{New York}
  \state{NY}
  \country{United States}
}

\author{Tengyu Xu}
\email{tengyuxu@meta.com}
\affiliation{%
  \institution{Meta Platforms Inc.}
  \city{Menlo Park}
  \state{CA}
  \country{United States}
}

\author{Boying Liu}
\email{boyingliu@meta.com}
\affiliation{%
  \institution{Meta Platforms Inc.}
  \city{Seattle}
  \state{WA}
  \country{United States}
}

\author{Wenbo Ren}
\email{wenboren@meta.com}
\affiliation{%
  \institution{Meta Platforms Inc.}
  \city{New York}
  \state{NY}
  \country{United States}
}

\author{Renkai Xiang}
\email{renkaixiang@meta.com}
\affiliation{%
  \institution{Meta Platforms Inc.}
  \city{Menlo Park}
  \state{CA}
  \country{United States}
}

\author{Songyi He}
\email{songyihe@meta.com}
\affiliation{%
  \institution{Meta Platforms Inc.}
  \city{Menlo Park}
  \state{CA}
  \country{United States}
}

\author{Haichuan Yang}
\email{hyang.ur@gmail.com}
\affiliation{%
  \institution{Meta Platforms Inc.}
  \city{Menlo Park}
  \state{CA}
  \country{United States}
}

\author{Serena Li}
\email{serenali@meta.com}
\affiliation{%
  \institution{Meta Platforms Inc.}
  \city{Menlo Park}
  \state{CA}
  \country{United States}
}

\author{Mingze Gao}
\email{gaomingze@meta.com}
\affiliation{%
  \institution{Meta Platforms Inc.}
  \city{Menlo Park}
  \state{CA}
  \country{United States}
}

\author{Yue Weng}
\email{yweng@meta.com}
\affiliation{%
  \institution{Meta Platforms Inc.}
  \city{Menlo Park}
  \state{CA}
  \country{United States}
}

\author{Ji Liu}
\email{ji.liu.uwisc@gmail.com}
\affiliation{%
  \institution{Meta Platforms Inc.}
  \city{Bellevue}
  \state{WA}
  \country{United States}
} 

\renewcommand{\shortauthors}{Xufeng Cai et al.}

\begin{abstract}
    Modern recommendation systems can be broadly divided into two key stages: the ranking stage, where the system predicts various user engagements (e.g., click-through rate, like rate, follow rate, watch time), and the value model stage, which aggregates these predictive scores through a function (e.g., a linear combination defined by a weight vector) to measure the value of each content by a single numerical score. Both stages play roughly equally important roles in real industrial systems; however, how to optimize the model weights for the second stage still lacks systematic study. This paper focuses on optimizing the second stage through auto-tuning technology.
    Although general auto-tuning systems and solutions - both from established production practices and open-source solutions - can address this problem, they typically require weeks or even months to identify a feasible solution. Such prolonged tuning processes are unacceptable in production environments for recommendation systems, as suboptimal value models can severely degrade user experience.
    An effective auto-tuning solution is required to identify a viable model within 2-3 days, rather than the extended timelines typically associated with existing approaches. 
    In this paper, we introduce a practical auto-tuning system named \hyperzero\ that addresses these time constraints while effectively solving the unique challenges inherent in modern recommendation systems. Moreover, this framework has the potential to be expanded to broader tuning tasks within recommendation systems.
\end{abstract}




\maketitle

\section{Introduction}\label{sec:intro}
Recommendation systems are widely deployed across various scenarios to facilitate user access to pertinent content~\citep{covington2016deep,gomez2015netflix,groh2012social}. In this context, numerous configuration hyperparameters play critical roles in achieving optimal performance and enhancing the overall user experience. To clarify this concept, we first provide a simplified view of how recommendation systems determine the order in which the content should be shown to users. The system usually consists of two components: 1) the ranking stage, and 2) the value model (VM) stage. 
In the ranking phase, a collection of models predicts the probabilities of different interactions between the user and the contents, such as $p(\mathrm{click})$, $p(\mathrm{share})$, and $p(\mathrm{follow})$. In the value model phase, these predictions are aggregated to determine the holistic value of each content to the user, which is then used to finally sort and present the best contents:
\begin{align*}
    \mathrm{score} = \theta_0 \ast p(\mathrm{click}) + \theta_1 \ast p(\mathrm{share}) + \theta_2 \ast p(\mathrm{follow}) + \cdots,
\end{align*}
where $\theta_0$, $\theta_1$, and $\theta_2$ are among the most common examples of hyperparameters in a recommendation system. To optimize these hyperparameters, the team needs to balance multiple aggregated online objectives such as total view counts, peak capacity cost, and user daily average usage time. These metrics often require days to calculate and are subject to considerable delay and noise. Traditional tuning approaches, even those refined by engineering teams over the years, typically rely on daily system feedback and require \emph{2-3 weeks} to find a reasonable balance.

\begin{figure*}[!t]
      \centering
      \includegraphics[width=0.85\textwidth]{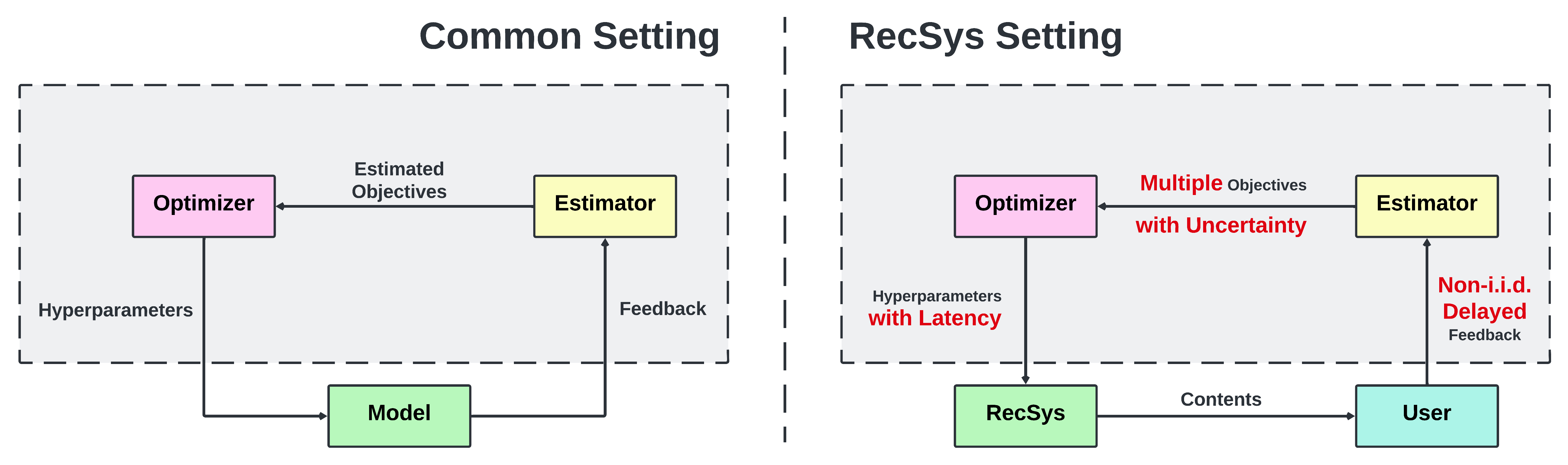}
      \caption{Hyperparameter tuning in common scenarios and recommendation systems respectively.}
      \label{fig:ill_naive_system}
\end{figure*}

In practice, however, these configuration parameters often need to be swiftly adapted to evolving business requirements, demanding high efficiency in tuning. For example, when a new interface is launched, there may be a need on the recommendation system to simultaneously promote longer videos to account for regionally nuanced user preferences while adhering to other business constraints. This necessitates an efficient solution for tuning configuration parameters that optimizes for multiple objectives under strict constraints within a short time frame, raising the following question:  
\begin{center}
\begin{tcolorbox}[enhanced,attach boxed title to top center={yshift=-3mm,yshifttext=-1mm},colback=gray!5!white,colframe=gray!75!black,colbacktitle=red!80!black,
  title=,fonttitle=\bfseries,
  boxed title style={size=small,colframe=red!50!black},width=0.48\textwidth, boxsep=1pt]
\emph{How do we reduce the tuning time from weeks to days?}
\end{tcolorbox}
\end{center}

To formalize this hyperparameter tuning problem in the context of recommendation systems, we first introduce a conceptual system that utilizes the tuning process, as illustrated on the left of Figure~\ref{fig:ill_naive_system}. This conceptual system consists of two main phases: the \emph{estimator} module, which estimates the objectives based on user feedback on the current hyperparameter, and the \emph{optimizer} module, which utilizes the estimation from the \emph{estimator} module and updates the hyperparameter. To address the efficiency concerns raised above, it is important that the tuning system leverages hourly system feedback. However, the conceptual workflow described faces substantial implementation challenges in practice. In particular, we illustrate the real-world system environment for hyperparameter tuning on the right of Figure~\ref{fig:ill_naive_system}, and discuss the emerging challenges as follows.

\begin{enumerate}[wide=0pt, leftmargin=\parindent]
    \item {\bf \emph{Estimator}: Non-i.i.d. hourly feedback.} Hourly system feedback data received by the prediction module exhibits high variance and is usually correlated over time, violating the independently and identically distributed (i.i.d.) condition required by most algorithms, which leads to a lack of performance guarantee. 
    \item {\bf \emph{Optimizer}: Multi-objective optimization.} Hyperparameter tuning typically involves multiple objectives on system metrics. However, each metric is a random variable with an unknown distribution, making direct estimation impossible. Furthermore, the  gradients or Hessians of these objectives are not available, making (stochastic) first- or second-order optimization algorithms inapplicable. Existing zeroth-order optimization methods, which require only function value evaluations, are also not directly applicable due to the presence of generic constraints and non-i.i.d. hourly data. 
    \item {\bf Delayed environment for system feedback.} There exist inherent system latencies associated with: 1) the hyperparameter change taking effect in the recommendation system, and 2) collecting user and system feedback after the hyperparameter change. These delays can span several hours and potentially offset the efficiency gains achieved by utilizing hourly data in the tuning process. 
\end{enumerate}
In this paper, we contribute to addressing the efficiency and capability challenges arising in hyperparameter tuning for recommendation systems. We overcome all aforementioned technical obstacles and propose an end-to-end system for \textbf{Hyper}parameter auto-tuning via \textbf{Zero}th-order optimization, namely \hyperzero.

\subsection{Contributions}
Our main contributions are summarized as follows.
\begin{itemize}[wide=0pt, leftmargin=\parindent]
    \item {\bf End-to-end auto-tuning with hourly feedback.} We develop an end-to-end hyperparameter auto-tuning system for recommendation systems. Our proposed \hyperzero\ framework is capable of optimizing multiple generic objectives and constraints and leverages \emph{hourly} system feedback to significantly reduce the iteration cycle from weeks to days. 
    \item \textbf{Semi-i.i.d. hourly signal.} To address the challenge of the non-i.i.d. hourly feedback, we propose a semi-i.i.d. (independently distributed) signal that tracks the metric delta between the test and base hyperparameters. This normalized signal is independently distributed on an hourly basis and characterizes the performance gain over the base hyperparameters. We further derive the mean and variance estimates for the delta signal using a high-order Taylor series approximation. 
    \item \textbf{Efficient zeroth-order constrained optimization.} \hyperzero\ accommodates multiple tuning objectives by solving a constrained optimization problem with generic objectives and constraints. We develop an efficient zeroth-order optimization framework that combines a Gaussian process estimator (GP) with Thompson sampling (TS) for arbitrary problem forms. The GP facilitates continuous estimation of random hourly signals with unknown distributions, while new candidates are proposed by solving the constrained problem using TS based on GP estimations. Additionally, we design a recurrent online scheduler for hyperparameter candidate proposals and revisits.
    \item {\bf Asynchronous parallel exploration.} We design and implement an asynchronous parallel scheme for hyperparameter sampling and exploration in \hyperzero. This parallelization strategy enhances efficiency by exploring multiple candidates simultaneously, while the asynchronous scheduler safeguards the tuning process against intermittent disruptions caused by inherent system delays and data waiting periods. 
\end{itemize}

\subsection{Related Work}\label{sec:relatedworks}
Hyperparameter tuning in recommendation systems has been a long-standing open problem, which has been studied only very recently~\citep{matuszyk2016comparative, galuzzi2020hyperparameter,montanari2022impact,dewancker2016bayesian, vanchinathan2014explore, moscati2023multiobjective}. However, most prior work has focused only on basic collaborative recommendation algorithms such as matrix factorization~\citep{matuszyk2016comparative, galuzzi2020hyperparameter,dewancker2016bayesian}, or deep learning-based systems~\citep{wu2023hyperparameter,zheng2023automl,joglekar2020neural}, and there is a lack of existing hyperparameter auto-tuning systems for large-scale modern recommendation systems.
On the other hand, Bayesian optimization (BO) methods have been extensively studied for generic hyperparameter tuning; see e.g.,~\cite{shahriari2015taking,frazier2018bayesian} for a comprehensive survey and~\cite{snoek2012practical, gpyopt2016, wu2016parallel, klein2017robo, emukit2018, kandasamy2020tuning,balandat2020botorch} for open-source BO frameworks. However, these state-of-the-art BO frameworks cannot be directly applied to our setting due to non-i.i.d. hourly feedback and constraints of arbitrary form.

\section{Problem Setting}\label{sec:prelim}
In this work we study the problem of hyperparameter tuning for recommendation systems. Specifically, we introduce our problem setting and define the following concepts. 

\noindent {\bf Hyperparameter $\vtheta$.\quad}
We denote the vector of configuration parameters to be tuned by $\vtheta \in \R^d$.
For example, when tuning the value model of a recommendation system, $\vtheta$ corresponds to the vector of scores assigned to each type of feedback event. 

\noindent {\bf System metric $X(\vtheta)$.\quad}
We use $X \in \R$ to denote the metric that characterizes user and system feedback in recommendation systems. Standard examples include 
\begin{itemize}[wide=0pt, leftmargin=\parindent]
    \item \emph{View count}: The total number of content items the user viewed on an app.
    \item \emph{Watch time}: The total time (in seconds or minutes) the user spent on an app.
\end{itemize}
These metrics are typically scheduled for collection and calculation on an hourly or daily basis in recommendation systems. In this work, we consider \emph{hourly} metrics for efficiency, which are random and depend on system hyperparameters, i.e.,\
\begin{align}\label{eq:metric}
    X = X(\vtheta; \vz), \quad \text{where } \vtheta \in \Theta,\; \vz \sim \gD_X
\end{align}
for some underlying unknown distribution $\gD_X$ that changes and evolves over hours. When the context is clear, we omit the notation $z$ and denote $X(\vtheta)$ for brevity. 

\noindent {\bf Constrained optimization formulation.\quad}
For any generic objective $f$ and constraints $g_i$ with respect to the system metrics $X_1(\vtheta), X_2(\vtheta), \dots$, hyperparameter tuning can be formulated as the following constrained optimization problem:
\begin{align}
    \max_{\vtheta} \;& \E\big[f\big(X_1(\vtheta), X_2(\vtheta), \dots\big)\big] \label{eq:objective} \\
    \text{ s.t. }\;& \E\big[g_i\big(X_1(\vtheta), X_2(\vtheta), \dots\big)\big] \geq c_i, \quad i = 1, 2, \dots, m. \label{eq:constr}
\end{align}
\begin{itemize}[wide=0pt, leftmargin=\parindent]
    \item Eq.~\eqref{eq:objective} stands for the average objective over a long time period and accounts for the randomness of metrics, such as the long-term gain in daily active users.
    \item Eq.~\eqref{eq:constr} imposes guardrail constraints on the hyperparameters to protect the ecosystem.
\end{itemize}
The functions $f$ and $g_i$ can take arbitrary forms, such as linear functions or even neural networks, and are typically determined by specific product scenarios. In general, their function values can be evaluated using the metric estimates for $X_1(\vtheta), X_2(\vtheta), \cdots$, but the gradient or Hessian of $f$ and $g_i$ with respect to $\vtheta$ is not available. The following example illustrates the optimization formulation in a real-world scenario for hyperparameter tuning.
\begin{example}\label{eg:constrained-opt}
Consider a scenario where we want to maximize the metric \emph{view counts} on an app, with a guardrail on its trade-off metric \emph{watch time} of more than $-0.1\%$. Combining this with further restrictions from the product side, such as limiting the delivery of content from sources not followed by the user to less than $10\%$, we have the following constrained problem 
\begin{align*}
        \max_{\vtheta} \;& \E[\text{\emph{view count}}(\vtheta)] \\
        \text{s.t. } \;& \E[\text{\emph{watch time}}(\vtheta)] \geq -0.1\%, \\ 
        \;& \E[\text{\emph{percentage of unfollowed contents}}(\vtheta)] \leq 10\%.
\end{align*}
\end{example}

\begin{figure}[!t]
  \centering
  \includegraphics[width=0.39\textwidth]{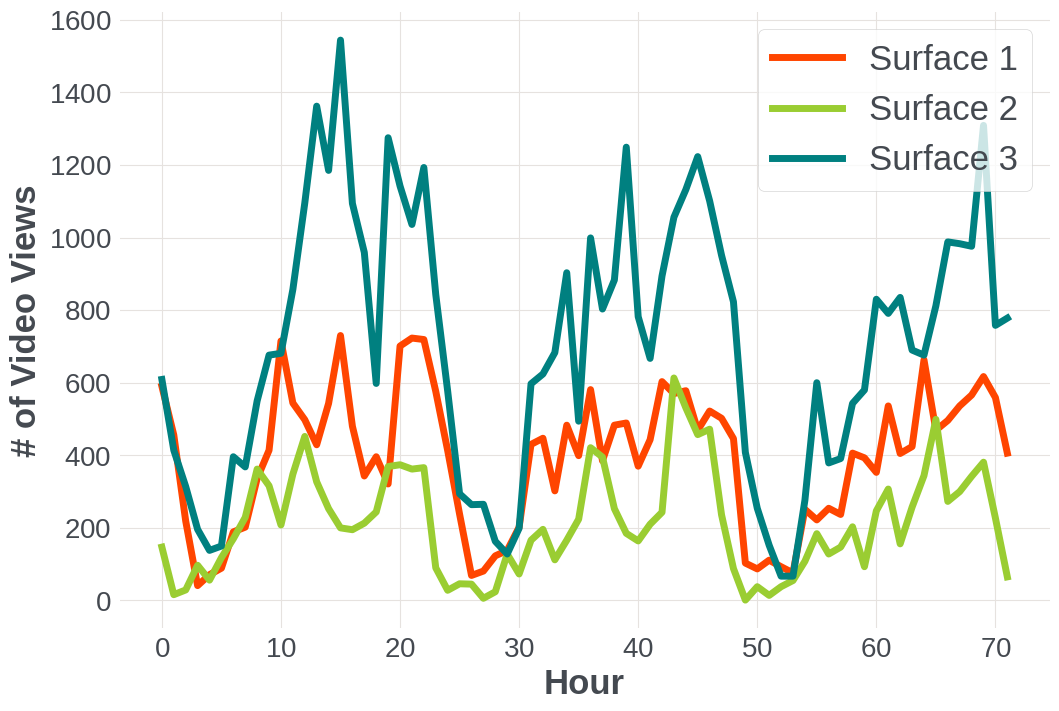}
  \caption{Hourly video view counts across different surfaces from an open-source dataset~\citep{gao2022kuairand}.}
  \label{fig:metricstrack}
\end{figure}
\section{\hyperzero: Auto-Tuning System}\label{sec:hyperzero}
In this section, we introduce our approaches for addressing the three challenges outlined in Section~\ref{sec:intro} and provide a detailed description of our \hyperzero\ auto-tuning system. We begin by introducing our data normalization approach and proposing an independently distributed signal from non-i.i.d. hourly metrics. We then develop a zeroth-order optimizer for solving the constrained problem defined in Eq.~\eqref{eq:objective}--\eqref{eq:constr}. To handle system latency and delayed feedback, we further present an asynchronous parallel scheme that allows the optimizer to explore multiple candidates simultaneously, enhancing both performance and efficiency.
Finally, we summarize and discuss the overall workflow of the \hyperzero\ system, combining all three components.

\subsection{Data Normalization}\label{sec:data-norm}
In this subsection, we address the first challenge arising from non-i.i.d. hourly metrics and propose a novel hourly signal based on the metric delta, as summarized below.
\begin{center}
\begin{tcolorbox}[enhanced,attach boxed title to top center={yshift=-3mm,yshifttext=-1mm},colback=gray!5!white,colframe=gray!75!black,colbacktitle=red!80!black,
  title=,fonttitle=\bfseries,
  boxed title style={size=small,colframe=red!50!black},width=0.48\textwidth, boxsep=1pt ]
\emph{{\bf Challenge 1}: Hourly metric is non-i.i.d. with correlation.\\
{\bf Solution}: Decorrelation via semi-i.i.d. $\Delta X(\vtheta) = \frac{X(\vtheta)}{X(\vtheta_0)} - 1$.} 
\end{tcolorbox}
\end{center}

We first provide further context on the challenge. To achieve our efficiency goal - convergence in days instead of weeks - \hyperzero\ operates on hourly system metrics. However, hourly data is noisier than daily data and is no longer independently distributed over hours. This is illustrated in Figure~\ref{fig:metricstrack}, where we plot the hourly trends of view counts across three different surfaces over $72$ consecutive hours using an open-source dataset~\citep{gao2022kuairand}. It can be observed that although a similar pattern repeats daily (every $24$ hours) for all surfaces, the hourly metric is non-i.i.d. and typically exhibits high variance and correlation over hours. Thus, it is hard to assess the performance of a given hyperparameter on an hourly basis, posing a significant challenge for hourly hyperparameter tuning.

To address this challenge, we propose a semi-i.i.d.  (independently distributed) hourly signal based on the ratio delta of the hourly metric readings between a test group and a control group, i.e.,\ 
\begin{align}\label{eq:delta-metric}
    \Delta X(\vtheta) := \frac{X(\vtheta)}{X(\vtheta_0)} - 1, \quad (\text{for a fixed base }\vtheta_0)
\end{align}
where $\vtheta_0$ is the fixed base hyperparameter assigned to the control group, representing either the default or the current system setup. Notably, $\Delta X(\vtheta)$ naturally characterizes the percentage metric gain of the test hyperparameter relative to the base hyperparameter. Our data normalization approach in Eq.~\eqref{eq:delta-metric} leverages the observation that user groups with different hyperparameter treatments exhibit a similar hourly fluctuation pattern, as illustrated in Figure~\ref{fig:metricstrack}. This indicates that metric deltas can be used to effectively decorrelate the hourly data. 

Note that although we obtain a semi-i.i.d. hourly signal, its distribution remains unknown. 
Hence, in order to facilitate zeroth-order optimization based on the estimates of $\Delta X$, we need to model its statistics - the mean $\mu(\Delta X; \vtheta)$ and the variance $\sigma^2(\Delta X; \vtheta)$ - to characterize the unknown distribution. However, it is technically nontrivial to estimate the mean and variance for the ratio of two random variables. To overcome this challenge, we approximate the statistics $\mu(\Delta X; \vtheta)$ and $\sigma^2(\Delta X; \vtheta)$ using a high-order Taylor expansion of the sample mean and variance of $X$ over $T$ rounds (hours), during which the hyperparameters $\vtheta$ and $\vtheta_0$ have been exposed. In particular, we use the following estimates: 
\begin{equation}\label{eq:delta-stat}
\begin{aligned}
    \mu(\Delta X; \vtheta) \coloneqq \;& \frac{1}{\sum_{t = 0}^T N_t}\sum_{t = 0}^T N_t\mu_t(\Delta X; \vtheta), \\
    \sigma^2(\Delta X; \vtheta) \coloneqq \;& \frac{1}{\big(\sum_{t = 0}^T N_t\big)^2}\sum_{t = 0}^T N_t^2 \sigma_t^2(\Delta X; \vtheta), \\
    \mu_t(\Delta X; \vtheta) \coloneqq \;& \frac{\mu_t(X; \vtheta)}{\mu_t(X; \vtheta_0)}  + \frac{\sigma_t^2(X; \vtheta_0)\mu_t(X; \vtheta)}{\mu^3_t(X; \vtheta_0)}, \\
    \sigma^2_t(\Delta X; \vtheta) \coloneqq \;& \frac{\mu_t^2(X; \vtheta_0)\sigma_t^2(X; \vtheta)/N'_t + \mu_t^2(X; \vtheta)\sigma_t^2(X; \vtheta_0)/N_t}{\mu^4_t(X; \vtheta_0)}, 
\end{aligned}
\end{equation} 
where $N_t$ and $N'_t$ denote the sizes (number of users) of the test group and the control group, respectively, and $\mu_t(X;\; \cdot)$ and $\sigma_t^2(X;\; \cdot)$ are the sample mean and variance of the metric readings for the user group at round $t$. A few notes are in order here. First, our estimates can be regarded as the weighted aggregation of hourly estimates $\mu_t(\Delta X; \vtheta)$ and $\sigma^2_t(\Delta X; \vtheta)$ over $T$ rounds, with more trust placed on hours with larger test groups. Further, the estimates $\mu(\Delta X; \vtheta)$ and $\sigma^2(\Delta X; \vtheta)$ can be calculated in a streaming manner in practice, i.e., they can be maintained over rounds and updated as new metric readings are collected.

Lastly, for notational convenience in later sections, we put together our individual estimates for possibly multiple metrics $\Delta \mX(\vtheta) = (\Delta X_1(\vtheta), \Delta X_2(\vtheta), \dots)^\top$ involved in the tuning task, i.e.,\ 
\begin{align*}
    \vmu_t(\vtheta) = \;& (\mu_t(\Delta X_1; \vtheta), \dots)^\top, \quad \vmu(\vtheta) = (\mu(\Delta X_1; \vtheta), \dots)^\top, \\ 
    \mSigma_t(\vtheta) = \;& \mathrm{diag}(\sigma^2_t(\Delta X_1; \vtheta), \dots), \quad \mSigma(\vtheta) = \mathrm{diag}(\sigma^2(\Delta X_1; \vtheta), \dots),
\end{align*}
where $\mathrm{diag}(\cdot)$ denotes a diagonal matrix. 

\begin{algorithm}[!t]
\caption{Zeroth Order Optimizer: GP + TS}\label{alg:optimizer}
\begin{algorithmic}[1]
\STATE \textbf{Inputs:} hyperparameters $\{\vtheta_{1}, \dots, \vtheta_{n}\}$, record of metric estimates $\gR = \{(\vmu_0(\vtheta_j), \mSigma_0(\vtheta_j), \vmu_1(\vtheta_j), \mSigma_1(\vtheta_j), \dots)\}_{j = 1}^n$, number of solution candidates $K$
\STATE Compute $\{(\vmu(\vtheta_j), \mSigma(\vtheta_j))\}_{j = 1}^n$ as in Eq.~\eqref{eq:delta-stat}
\FOR{$k = 1, \dots, K$}
\STATE Estimate $\Delta \mX(\vtheta_{j}) \sim \mathcal{N}\big(\vmu(\vtheta_{j}), \mSigma(\vtheta_{j})\big)$ for all $j = 1, \dots, n$
\STATE Compute $f(\Delta \mX(\vtheta_{j}))$ and $g_i(\Delta \mX(\vtheta_{j}))$ for all $j = 1, \dots, n$
\STATE Compute $\Theta = \big\{\vtheta_j \mid g_i(\Delta \mX(\vtheta_{j})) \geq c_i, i = 1, \dots, m \big\}$
\STATE Pick $\vtheta_{*}^{(k)} = \argmax_{\vtheta_j \in \Theta} f(\Delta \mX(\vtheta_j))$ 
\ENDFOR
\STATE Return $\{\vtheta_*^{(k)}\}_{k = 1}^K$ and $\{(\vmu(\vtheta_j), \mSigma(\vtheta_j))\}_{j = 1}^n$
\end{algorithmic}
\end{algorithm}

\subsection{Zeroth-Order Optimization (GP + TS) for Multiple Constraints}
We now discuss how to solve the constrained optimization problem as in Eq.~\eqref{eq:objective}--\eqref{eq:constr}.  Incorporating metric deltas, we aim to solve 
\begin{equation}\label{eq:optimizationproblem}
\begin{aligned}
    \max_{\vtheta} \;& \E\big[f\big(\Delta \mX(\vtheta)\big)\big] \\
    s.t. \;& \E\big[g_i\big(\Delta \mX(\vtheta)\big)\big] \geq c_i, \quad i = 1, 2, \dots, m.
\end{aligned}
\end{equation}
However, the problem in Eq.~\eqref{eq:optimizationproblem} poses two main challenges in the context of recommendation systems, as summarized below. 
\begin{center}
\begin{tcolorbox}[enhanced,attach boxed title to top center={yshift=-3mm,yshifttext=-1mm},colback=gray!5!white,colframe=gray!75!black,colbacktitle=red!80!black,
  title=,fonttitle=\bfseries,
  boxed title style={size=small,colframe=red!50!black},width=0.48\textwidth, boxsep=1pt ]
\emph{{\bf Challenge 2(a)}: Random $\Delta \mX(\vtheta)$ with unknown distribution.\\
{\bf Solution}: Continuous Gaussian process (GP) estimation.} \\
\emph{{\bf Challenge 2(b)}: $\E[f(\cdot)]$, $\E[g_i(\cdot)]$'s gradients unavailable.\\
{\bf Solution}: Thompson sampling (TS)-based optimization.}
\end{tcolorbox}
\end{center}
For those challenges that have particularly arisen in recommendation system settings, existing optimization approaches are not applicable. In particular, the absence of gradient information precludes the use of gradient-based algorithms, so one needs to assort to certain zeroth-order optimization techniques. However, standard zeroth-order approaches typically rely on certain inductive bias regarding the posterior distribution across consecutive iterations, which is nontrivial to establish for generic objectives and constraints in recommendation system settings.

We propose a custom zeroth-order optimization framework that combines Gaussian process (GP) modeling with Thompson sampling (TS), effectively addressing both challenges. We summarize each round of the framework in Algorithm 1, while the overall algorithm is presented in Algorithm~\ref{alg:overall-optimizer}. In the following, we first introduce the two main components - GP and TS - each of which tackles a specific challenge, and then discuss the overall algorithm framework in detail.

\begin{figure*}[!t]
      \centering
      \includegraphics[width=0.7\textwidth]{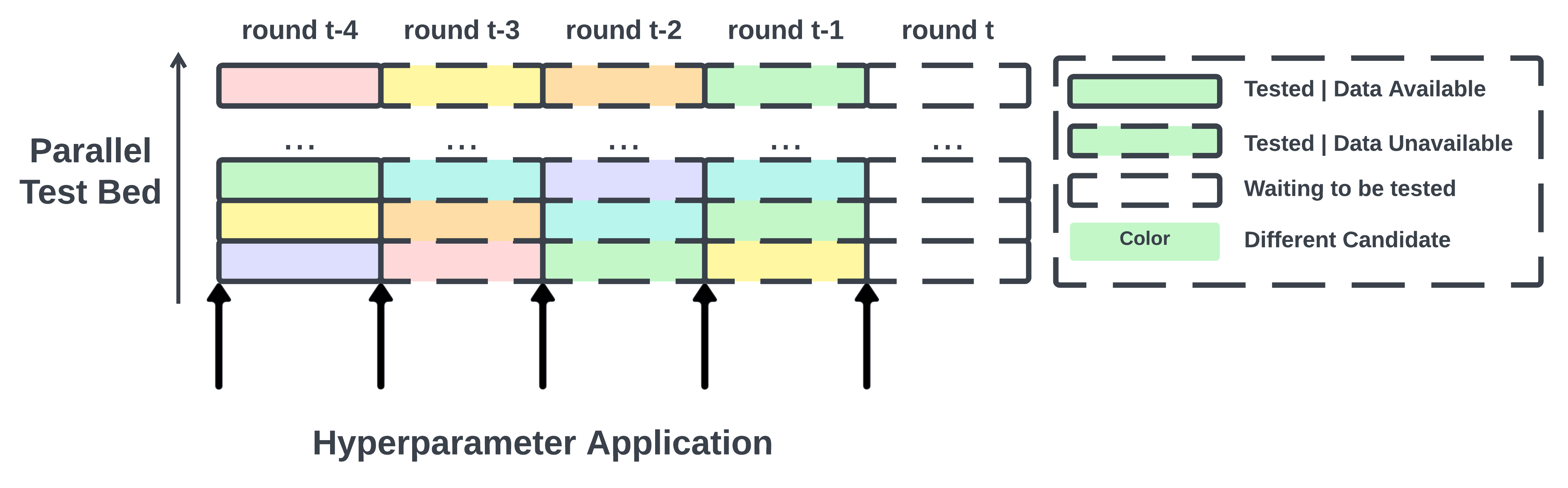}
      \caption{Workflow of asynchronous parallelization in \hyperzero.} 
      \label{fig:async}
\end{figure*}

\subsubsection{Gaussian process estimation for Challenge 2(a).}
To start with, we note that although $f$ and $g_i$ can be evaluated once the metrics are estimated, particularly in scenarios of recommendation systems (see e.g.,\ Example~\ref{eg:constrained-opt}), the metric deltas $\Delta \mX$ remain random variables with unknown distributions.
To deal with this issue, we use Gaussian process (GP) for the continuous estimation of the unknown distribution of each metric delta, i.e.,\ 
\begin{align}
    \Delta X_i(\vtheta) \sim \gN\big(\mu(\Delta X_i; \vtheta), \sigma^2(\Delta X_i; \vtheta)\big), \quad i = 1, 2, \dots, \tag{GP}
\end{align}
where $\mu(\Delta X_i; \vtheta), \sigma^2(\Delta X_i; \vtheta)$ are the mean and variance estimates of $\Delta X_i$ at the current round, as in Eq.~\eqref{eq:delta-stat}. The detailed steps of GP estimation are listed in Lines 2-5 of Algorithm~\ref{alg:optimizer}, where a record $\gR$ of the hourly metric estimates for candidates $\{\vtheta_j\}_{j = 1}^n$ are maintained for the GP estimation process. The GP estimates $\Delta \mX(\vtheta_j)$ are then utilized to evaluate the objective and constraint values.
Here, we assume the prior distribution of the metric delta to be Gaussian, following Bayesian optimization approaches where Gaussian process regression is commonly employed.
More advanced techniques, such as neural networks, could also be used to model the unknown distribution, which we leave for future exploration.

\subsubsection{Thompson sampling for Challenge 2(b)}
For generic objective and constraint functions, their gradient or Hessian with respect to $\vtheta$ is usually inaccessible, so standard first-order or second-order methods in optimization literature do not apply in this setting. 
To address this challenge on the optimization side, we adopt Thompson sampling (TS) to pick candidates that satisfy the constraints and maximize the objective, as stated in Lines 6--7 of Algorithm~\ref{alg:optimizer}:
\begin{align}
    \vtheta_{*}= \argmax_{\vtheta} f(\Delta \mX(\vtheta)) \; \text{s.t. }\; g_i(\Delta \mX(\vtheta)) \geq c_i, \; i = 1, \dots, m. \tag{TS}
\end{align}
Since the objective and constraint values for TS are computed based on randomly sampled metric deltas, Algorithm~\ref{alg:optimizer} repeats the TS process $K$ times to generate multiple solution candidates. 

A few remarks are in order here. First, an astute reader may notice that there could be duplicate candidates. This is actually desirable, as more attention will be drawn to the more promising candidates in $\{\vtheta_j\}_{j = 1}^n$. 
Second, we note that whether a candidate satisfies all the constraints (Line $6$ of Algorithm~\ref{alg:optimizer}) can be checked efficiently as the constrained problem in Eq.~\eqref{eq:optimizationproblem} is usually low-dimensional (with fewer than $10$ hyperparameters at a time). 

For high-dimensional problems, log-barrier methods or other constraint-penalized formulations can be incorporated into \hyperzero\ to solve the constrained problem more efficiently.

\begin{algorithm}[!t]
\caption{Algorithm Framework of \hyperzero}\label{alg:overall-optimizer}
\begin{algorithmic}[1]
\STATE {\bf Input:} total number of rounds $T$, proposal probability $p$, number of solution candidates $K$, number of samples for proposal $N$
\STATE {\bf Initialize:} Hyperparameter candidate bucket $\gB$, metric estimate record $\gR = \{(\vmu_0(\vtheta), \mSigma_0(\vtheta))\}_{\vtheta \in \gB}$
\FOR{t = 1, \dots, T}
\STATE Log $\{(\vmu_t(\vtheta), \mSigma_t(\vtheta))\}_{\vtheta \in \gB}$ to $\gR$
\STATE Compute $(\gA, \gR') = \mathtt{Algorithm~\ref{alg:optimizer}}(\gB, \gR, K)$
\STATE $\mathtt{GP} = $ Gaussian Process Regression on $(\gB, \gR')$ 
\STATE $\gA' = \mathtt{Algorithm~\ref{alg:proposal}}(N, \mathtt{GP})$ with probability $p$ otherwise $\emptyset$
\STATE Submit $\gA \cup \gA'$ for online tests
\STATE Update $\gB = \gB \cup \gA'$
\ENDFOR
\end{algorithmic}
\end{algorithm}

\subsubsection{Overall algorithm framework.}
We now discuss the overall algorithm framework of \hyperzero, as summarized in Algorithm~\ref{alg:overall-optimizer}. In the beginning, a bucket of hyperparameter candidates is initialized either through a prescribed choice or by uniformly grid-sampling the hyperparameter space. Typically, in practice, an initial bucket size of $100$ candidates is sufficient. For each candidate, we maintain a record $\gR$ of all its historical hourly metric estimates. At each round, we collect the latest hourly metrics readings from system feedback and log the corresponding hourly estimates onto $\gR$. Then we call Algorithm~\ref{alg:optimizer} to solve the constrained problem in Eq.~\eqref{eq:optimizationproblem} and pick the solution candidates for online tests to collect metric readings.

To better explore the hyperparameter space and reduce the dependence on the initialization, we further introduce a subroutine of proposing new candidates in each round. The detailed proposal procedure is summarized in Algorithm~\ref{alg:proposal}, and \hyperzero\ proposes and adds new candidates to the bucket with probability $p$ (usually set to $1$) in each round. To generate a new candidate, Algorithm~\ref{alg:proposal} randomly samples $N$ points from the space and adopts the idea of TS, in contrast to less efficient Monte Carlo sampling. To obtain metric estimates for a newly sampled point, we use Gaussian process regression $\mathtt{GP}$ to interpolate on the current bucket of hyperparameter candidates and their latest aggregated estimates $\gR'$, which provides better metric estimates for Algorithm~\ref{alg:proposal}.

Picking the solution candidates $\gA$ using Algorithm~\ref{alg:optimizer} for online tests is necessary for two reasons. First, the online traffic for A/B tests is typically limited, while we may have hundreds of candidates to test. Hence, we can only send a subset of the candidates for online tests and collect their metric readings. On the other hand, we want more online traffic to be directed toward the more promising candidates. The repeated TS process in Algorithm~\ref{alg:optimizer} would add promising candidates in $\gA$ with duplicates. Since each candidate receives the same amount of online exposure, duplicated candidates in $\gA$ automatically receive more traffic for online tests.

\begin{algorithm}[!t]
\caption{Candidate Proposal}\label{alg:proposal}
\begin{algorithmic}[1]
\STATE \textbf{Inputs:} number of samples $N$, distribution surrogate $\mathtt{GP}$
\STATE Randomly sample $\{\vtheta_j\}_{j = 1}^N$ in the hyperparameter space
\FOR{$j = 1, \dots, N$}
\STATE Compute $\big(\vmu(\vtheta_j), \mSigma(\vtheta_j)\big) = \mathtt{GP}(\vtheta_j)$ 
\STATE Estimate $\Delta \mX(\vtheta_{j}) \sim \mathcal{N}\big(\vmu(\vtheta_{j}), \mSigma(\vtheta_{j})\big)$
\STATE Compute $f(\Delta \mX(\vtheta_{j}))$ and $g_i(\Delta \mX(\vtheta_{j}))$
\ENDFOR
\STATE Compute $\Theta = \big\{\vtheta_j \mid g_i(\Delta \mX(\vtheta_{j})) \geq c_i, i = 1, \dots, m\big\}$
\STATE Pick $\vtheta_* = \argmax_{\vtheta_j \in \Theta} f(\Delta \mX(\vtheta_{j}))$ 
\STATE Return $\vtheta_*$ 
\end{algorithmic}
\end{algorithm}

\begin{figure*}[!t]
  \centering
\includegraphics[height=0.4\textwidth]{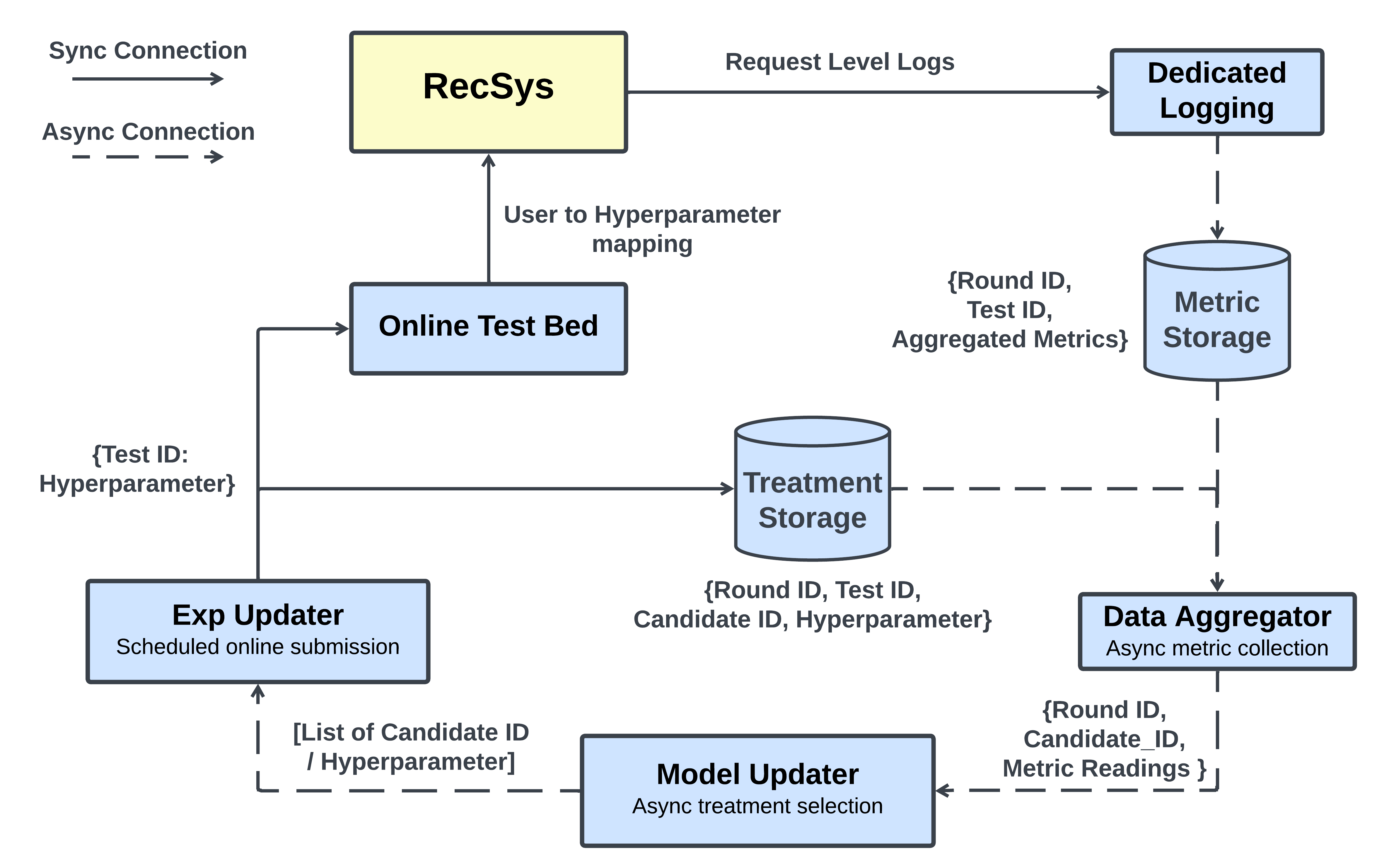}\label{fig:high-level}
  \caption{System design of \hyperzero\ auto-tuning system.}
  \label{fig:hyperzero}
\end{figure*}

\subsection{Asynchronous parallel exploration}
We now introduce the final component of \hyperzero\ - asynchronous parallelization - to implement our algorithm framework in a real-world recommendation system, addressing system latency for hourly metric readings.
\begin{center}
\begin{tcolorbox}[enhanced,attach boxed title to top center={yshift=-3mm,yshifttext=-1mm},colback=gray!5!white,colframe=gray!75!black,colbacktitle=red!80!black,
  title=,fonttitle=\bfseries,
  boxed title style={size=small,colframe=red!50!black},width=0.48\textwidth, boxsep=1pt ]
\emph{{\bf Challenge 3}: System latency for hourly metric readings.\\
{\bf Solution}: Asynchronous parallel exploration.}
\end{tcolorbox}
\end{center}

\subsubsection{Parallelization}
As stated in Algorithm~\ref{alg:overall-optimizer}, \hyperzero\ maintains and explores a bucket of hyperparameter candidates in each iteration. To improve the efficiency in terms of wall-clock time, these candidates are explored \emph{in parallel}  to collect hourly metric readings. This parallelization approach allows for more testing results within the same time period, compared to exploring one candidate at a time with all online traffic.

\subsubsection{Asynchronous execution}
Further, parallelization faces a bottleneck caused by inherent system latency of hyperparameter configuration changes, and system feedback for all candidates is often not received simultaneously by the tuning system. To resolve this issue, \hyperzero\ is built as an \emph{asynchronous} parallel loop at the infrastructure level, as illustrated in a typical running scenario in Figure~\ref{fig:async}. In particular, at a certain starting point (e.g., 0 minutes, 0 seconds of an hour), the system will apply multiple hyperparameter candidates in parallel on the test bed (usually a group of online A/B tests), with the understanding that feedback from the most recent rounds may or may not be available. As shown in Figure~\ref{fig:async}, the feedback data is delayed by $3$ rounds, and at each round, \hyperzero\ makes the hyperparameter application decisions based on all available data at that point, i.e.,\ updating the record $\gR$ of metric estimates and proceeding with Algorithm~\ref{alg:optimizer}. For each candidate, its estimates in Eq.~\eqref{eq:delta-stat} are aggregated only over the rounds when the estimates are available in $\gR$, as Eq.~\eqref{eq:delta-stat} does not require consecutive hourly estimates. In practice, even with a delay spanning over $6$ rounds, the convergence speed of \hyperzero\ remains almost the same during a tuning job over $3$ days.

\subsection{Overall system design}
To conclude this section, we now present the system design of \hyperzero\ in Figure~\ref{fig:hyperzero}, combining all the components introduced in the previous subsections, where we implement our algorithm framework of \hyperzero\ as outlined in Algorithm~\ref{alg:overall-optimizer} within a real system environment. In each round, the $\mathtt{data \; aggregator}$ asynchronously collects user feedback and system metrics. The $\mathtt{model \; updater}$ logs the most recent metric estimates and follows Algorithm~\ref{alg:overall-optimizer} to solve the constrained optimization problem in Eq.~\eqref{eq:optimizationproblem} based on the available data at the current round. The candidates $\gA \cup \gA'$ picked by the $\mathtt{model \; updater}$ are then sent to the $\mathtt{exp \; updater}$, which submits the candidates to the online testbed for application in the recommendation system. In addition to these algorithm modules, the hyperparameter configurations and their metric readings and estimates are stored in the $\mathtt{hyperparameter \; storage}$ and $\mathtt{metric \; storage}$, respectively, in the format specified in Figure~\ref{fig:hyperzero}.

\section{Experiments}\label{sec:experiment}
In this section, we conduct numerical experiments to evaluate our proposed \hyperzero\ framework, based on the following questions:
\begin{itemize}[wide=0pt, leftmargin=\parindent]
    \item {\bf Q1 (ablation study)}: Is each algorithm component of \hyperzero\ essential for its performance with hourly metrics?
    \item {\bf Q2 (performance comparison)}: Does \hyperzero\ show better performance than existing tuning approaches?
\end{itemize}
Note that, unlike many other papers, this paper essentially introduces a full system rather than a single algorithm, which makes it hard to perform an apple-to-apple comparison. Therefore, the experiments will be mainly used to validate each component of this system.
We first test the \hyperzero\ system in a synthetic environment for ablation studies and present a preliminary comparison with a general Bayesian optimization-based hyperparameter tuning framework. The main reason for using synthetic data is that there are no open-source datasets where we can explicitly evaluate the metrics given certain hyperparameters. To justify the performance in real-world scenarios, we then present the evaluation results, including further ablation studies and comparisons with a leading open-source Bayesian optimization framework, by deploying \hyperzero\ in a production environment. 

\subsection{Experiments on synthetic data}\label{sec:exp-synthetic}
In this subsection, we consider a synthetic environment defined in Sec.~\ref{sec:exp-setting-synthetic}. In this environment, $1$) we first conduct ablation studies for \hyperzero\ for {\bf Q1}, where we remove one main design component from \hyperzero\ at a time. In particular, we consider two variants of \hyperzero: one adopts synchronous parallelization and proceeds only after all batches of feedback have been received, while the other directly uses the hourly metrics without data normalization; $2$) For {\bf Q2}, we compare \hyperzero\ with a built-in Bayesian optimizer from scikit-optimize~\cite{tim_head_2018_1207017} adapted to this setting. For both cases, the performance is evaluated based on the following two measures: the objective gain over the base hyperparameter and the constraint violation defined by
\begin{align}\label{eq:eval-metric}
    \mathrm{Gain}(\vtheta) := \frac{f(\vtheta)}{f(\vtheta_0)} - 1, \; \mathrm{Violation}(\vtheta) := \max\{c - g(\vtheta), 0\}, 
\end{align}
where a larger objective gain is preferable and a smaller or zero constraint violation is desirable.

\subsubsection{Experiment setting.}\label{sec:exp-setting-synthetic}
Motivated by the hyperparameter tuning scenario mentioned in Example~\ref{eg:constrained-opt}, we consider the following constrained optimization problem: 
\begin{equation}\label{eq:exp-prob}
\begin{aligned}
    \max_{\vtheta\in [0, 1]^2} \;& \Big\{f(\vtheta) := \vw_1 \mathbbm{E}\left[X_1(\vtheta; t)\right] + \vw_2 \mathbbm{E}\left[X_2(\vtheta; t)\right]\Big\} \\
    s.t. \;& g(\vtheta) := \vw_3\mathbbm{E}\left[X_1(\vtheta; t)\right] + \vw_4\mathbbm{E}\left[X_2(\vtheta; t)\right] \ge c 
\end{aligned}
\end{equation}
where $\vtheta$ is the hyperparameter to optimize, $t$ represents the hourly time step, and $X_1(\vtheta; t)$, $X_2(\vtheta; t)$ simulate two hourly metrics with uncertainty following Gaussian distributions. The base hyperparameter $\vtheta_0$ is randomly initialized and then fixed for the duration of the experiment. We further construct a synthetic environment over $T$ time steps (hours) to simulate the recommendation system with two sources of latency. First, the metric reading is delayed by $\tau$ time steps after the new $\vtheta$ is dispatched. Second, the system takes an additional $\xi$ time steps to receive the response metrics, where $\xi$ is a random Gaussian variable.
Details on the experiment setup and parameter choices are provided in Appendix~\ref{app:exp_detail}. 

\subsubsection{Experiment results.} 
We now report the results of two experiments on synthetic data.

\begin{figure}[!t]
    \centering
    \hspace*{\fill}
    \subfigure[Objective gain]{\includegraphics[width=0.23\textwidth]{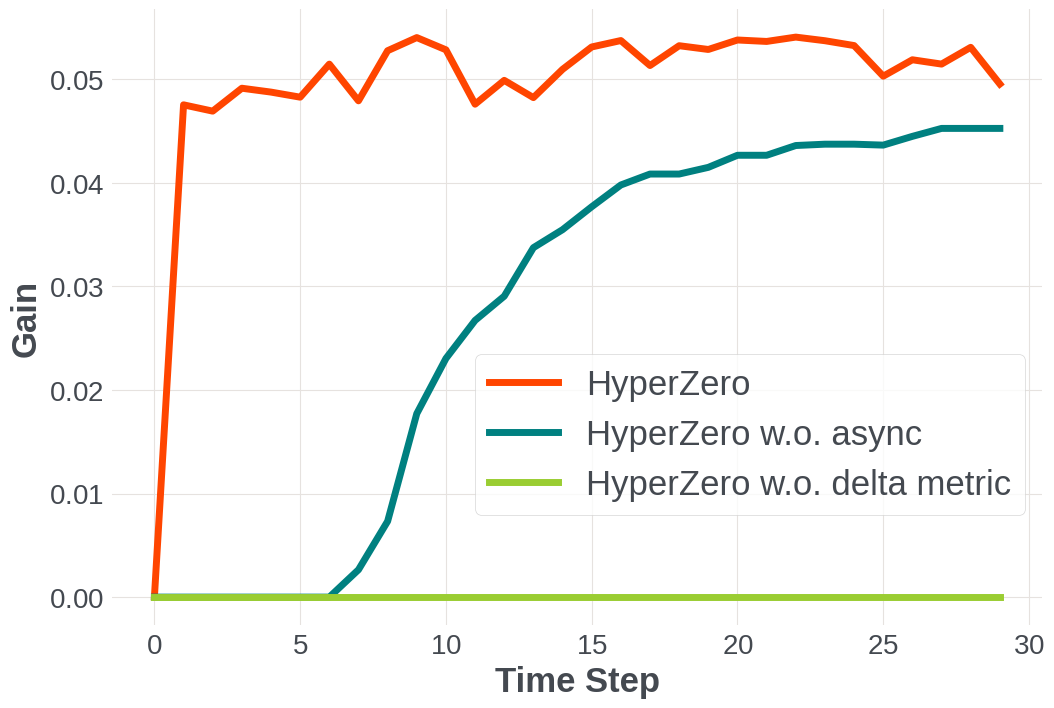}\label{fig:gain}}
    \hfill
    \subfigure[Constraint violation]{\includegraphics[width=0.23\textwidth]{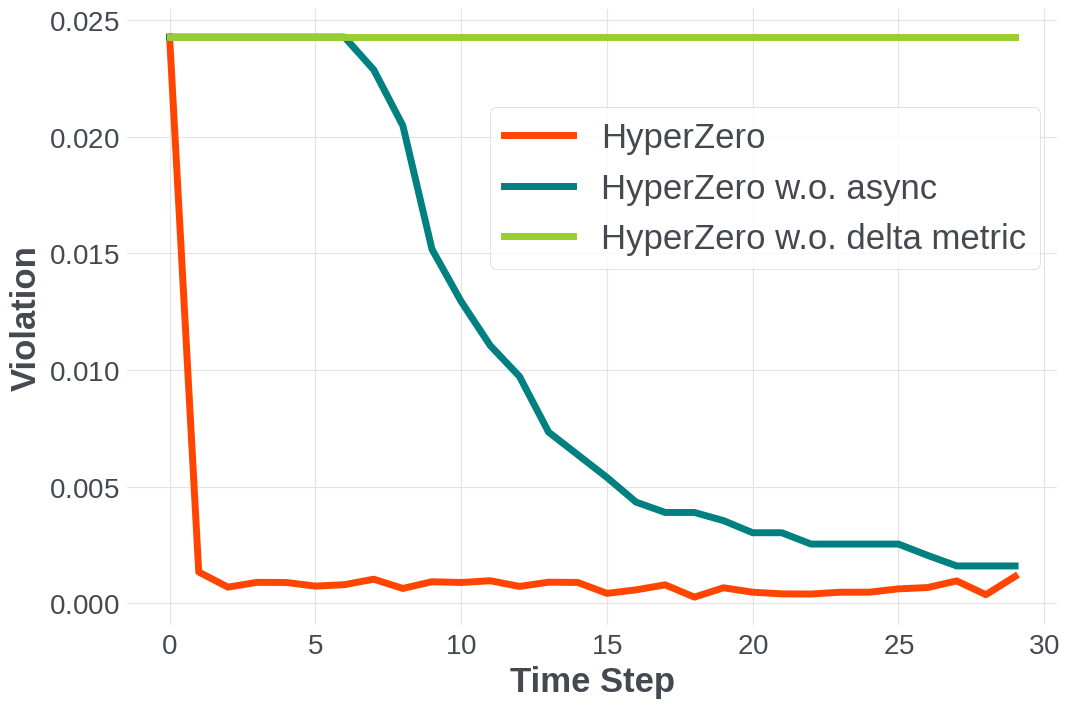}\label{fig:violation}}
    \hspace*{\fill}
    \caption{Ablation study of \hyperzero\ on synthetic data.}
    \label{fig:ablation-synthetic}
\end{figure}

\paragraph{Ablation study}
We plot the objective gain and constraint violation for \hyperzero\ and its ablated variants in Figure~\ref{fig:ablation-synthetic}, with the results averaged over $50$ runs using different random seeds.
In response to {\bf Q1}, we observe that $1$) \hyperzero\ demonstrates a much faster learning speed for the objective gain and better control of the constraint violation compared to the other two variants; $2$) The delta metric is critical, as the algorithm cannot learn anything from the environment if it directly uses the hourly metrics; $3$) The synchronous variant learns from the environment approximately $50\%$ slower than \hyperzero, validating the efficiency improvement through asynchronous parallelization. 

\paragraph{Comparison with Bayesian optimizer}
Since the Bayesian optimizer (BO) in scikit-optimize~\cite{tim_head_2018_1207017} only accepts unconstrained formulations, we compare \hyperzero\ with two BO variants adapted to our setting: one that optimizes only the objective (unconstrained) and the other that optimizes a constraint-penalized objective (penalized). The regularization parameter of the penalized BO variant is manually tuned to be optimal. Ignoring the gains of \hyperzero\ from data normalization, we also test both BO variants on normalized delta metrics; otherwise, as illustrated in Figure~\ref{fig:ablation-synthetic}, the solver cannot learn effectively from the environment. Other implementation details for BO follow the solver's default settings.

We report the comparison results after $30$ time steps over $50$ runs with different random seeds in Table~\ref{tab:bo-synthetic}. The results show that \hyperzero\ outperforms the BO methods in both measures, and \hyperzero\ even achieves a higher objective gain while solving a constrained optimization problem. Furthermore, the hard constraints implemented in \hyperzero\ ensure better guardrail feasibility, while the penalized BO still suffers significant guardrail violations, even with a large regularization. These results provide an affirmative answer to {\bf Q2}, demonstrating that \hyperzero\ has better exploration efficiency and is capable of finding solutions that are more globally optimal than those identified by BO. 

\begin{table}[t]
  \caption{Comparison between \hyperzero\ and Bayesian optimization on delta metrics on synthetic data.}
  \label{tab:bo-synthetic}
  \begin{tabular}{lcc}
    \toprule
     & Gain ($\%$) & Violation \\
    \midrule
    Unconstrained BO & $2.128 \pm 0.856$ & $0.592 \pm 0.0051$ \\
    \addlinespace
    Penalized BO & $2.080 \pm 0.760$ & $0.592 \pm 0.0044$ \\
    \addlinespace
    \hyperzero\ & $\mathbf{4.951 \pm 0.205}$ & $\mathbf{0.001 \pm 0.0004}$ \\
  \bottomrule
\end{tabular}
\end{table}

\subsection{Experiments in industrial settings}\label{sec:experiment-online}

For {\bf Q1}, we provide a more fine-grained case study on the ablated variant of \hyperzero. For {\bf Q2}, we compare the optimizer component of \hyperzero\ with an industry-leading open-source Bayesian optimizer, complementing the comparison in the previous subsection. In both experiments, we consider the tuning task for a set of VM hyperparameters in the recommendation system and report the \emph{performance gain} defined in Eq.~\eqref{eq:eval-metric} for the task's target metric. 

\subsubsection{Ablation study}
\begin{figure}[!t]
    \centering
    \includegraphics[width=0.39\textwidth]{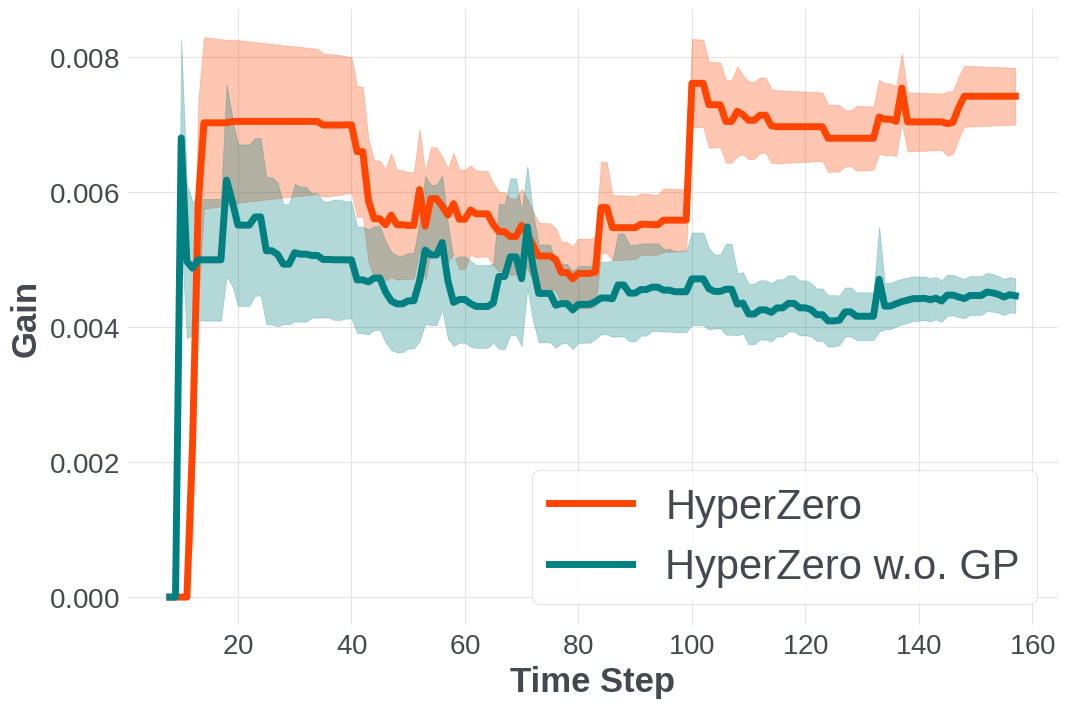}
    \caption{Ablation study of candidate proposal of \hyperzero\ in industrial settings.}
    \label{fig:online-ablation}
\end{figure}

We now conduct a further case study on the more fine-grained algorithm ingredients of \hyperzero. In particular, we study whether incorporating candidate proposals ($p \neq 0$) as in Algorithm~\ref{alg:proposal} further improves metric gains.
We show that exploration with candidate proposals ($p \neq 0$ in Algorithm~\ref{alg:overall-optimizer}) leads to better performance gains. Specifically, we consider an unconstrained tuning task aimed at increasing view counts on an app and compare \hyperzero\ with its variant that does not use candidate proposals ($p = 0$) for this task. We plot the performance gains over the base setup versus hours for $3$ days in Figure~\ref{fig:online-ablation}. The results show that \hyperzero\ achieves double the performance gain compared to \hyperzero\ without candidate proposals. This confirms that exploring new candidates using Algorithm~\ref{alg:proposal} is beneficial to improve the initial prescribed candidates. 

\subsubsection{Comparison with Bayesian optimizer} 
We compare the optimizer component of \hyperzero\ with the Bayesian optimizer Botorch~\citep{balandat2020botorch}, one of the most popular open-source auto-tuning tools. Since Botorch is not specifically designed for or directly applicable to recommendation systems, we integrate it into our \hyperzero\ framework with the data normalization component. This comparison aims to validate that the optimizer in \hyperzero\ is at least on par with the mainstream open-source optimizers. Both methods are applied to tune $4$ VM hyperparameters in the same production application. Each hyperparameter candidate is exposed to millions of users via the \hyperzero\ system before being put to online testing for comparison. For a fairer comparison, we consider the tuning task of increasing view counts without guardrail constraints, disregarding differences in how the two methods enforce constraints. 

\begin{figure}[!t]
    \centering
    \hspace*{\fill}
    \subfigure[Performance gain of \hyperzero.]{\includegraphics[width=0.23\textwidth]{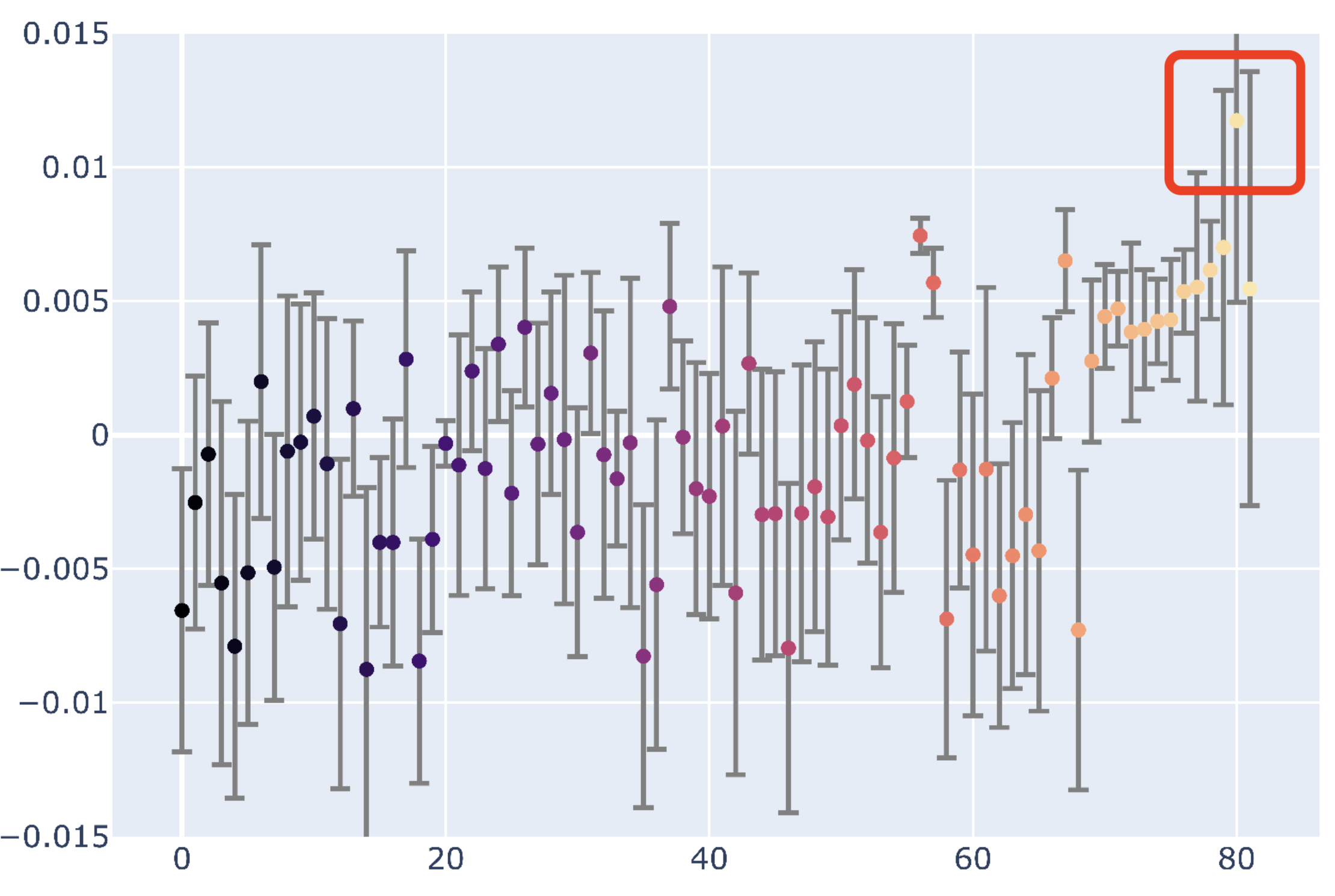}\label{fig:online-hyperzero}}
    \hfill
    \subfigure[Performance gain of Botorch.]{\includegraphics[width=0.23\textwidth]{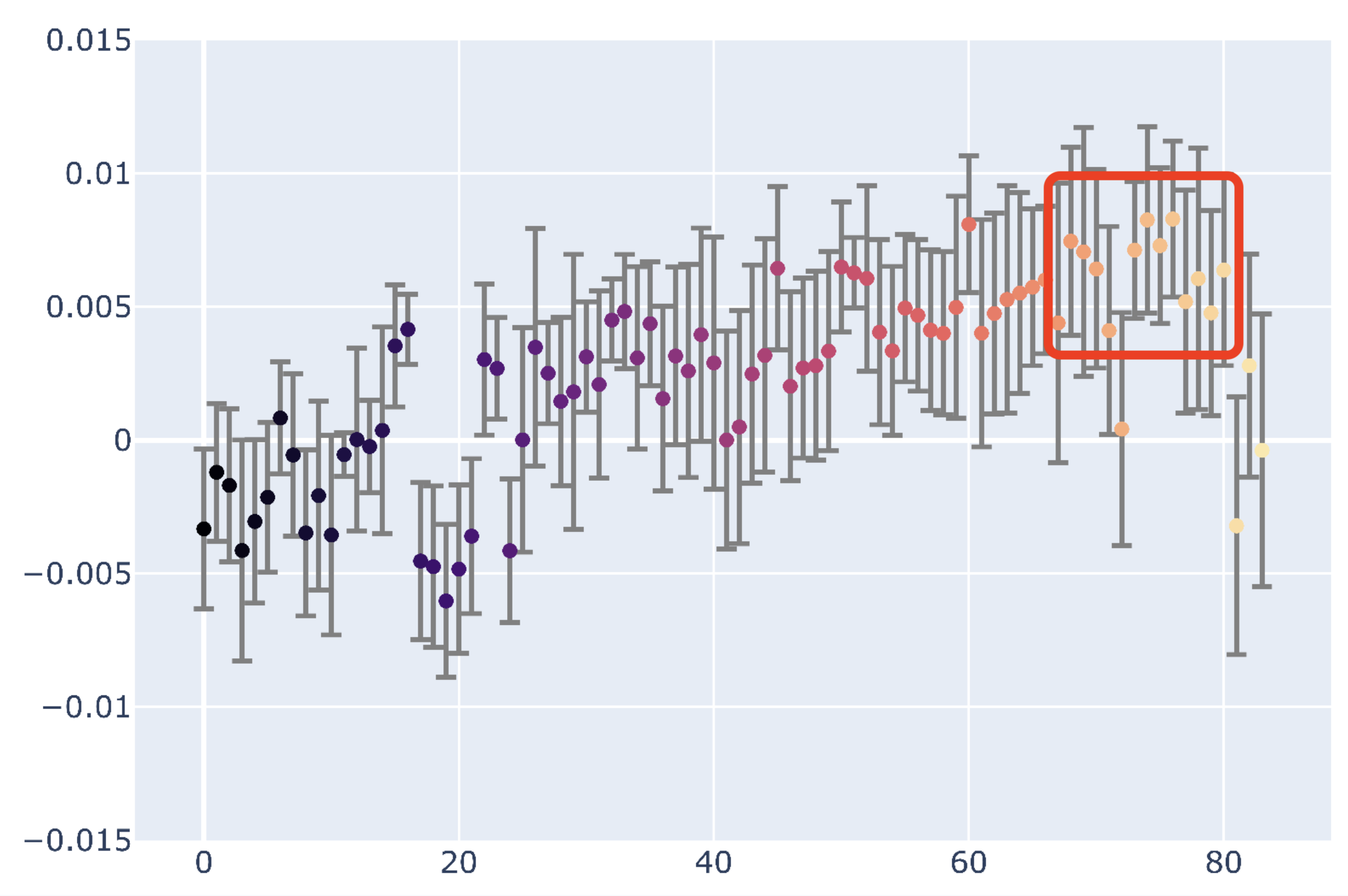}\label{fig:online-bo}}
    \hspace*{\fill}
    \caption{Comparison between \hyperzero\ and Botorch in industrial settings: gains $1.2\%$ (\hyperzero) vs. $0.8\%$ (Botorch) in terms of mean value.}
    \label{fig:ablation-online}
\end{figure}

We plot the hourly trends of view count gains for each method over $3$ days in Figure~\ref{fig:ablation-online}, where the color indicates the recency of each candidate's proposal by the optimizer. The comparison criteria are the performance of the best candidates from the two methods in terms of mean value and lower confidence bound (LCB). 
As highlighted in the figure, the best results of \hyperzero\ in Figure~\ref{fig:online-hyperzero} outperform Botorch's in Figure~\ref{fig:online-bo} on both criteria: $1.2\%$ vs. $0.8\%$ (mean value) and $0.7\%$ vs. $0.5\%$ (LCB). It is worth noting that the highlighted point in Figure~\ref{fig:online-hyperzero} is also the best candidate in terms of the upper confidence bound (UCB). This point will undergo more online tests as long as the experiment continues, making it a high-confidence candidate if the same gain is observed in subsequent exploration. Additionally, we observe that Botorch is more exploitation-dominant. Hence, most candidates in Figure~\ref{fig:online-bo} perform well, but identifying significantly better candidates is inefficient, as evidenced by the curve drop at the end of Figure~\ref{fig:online-bo}. In contrast, \hyperzero\ facilitates more efficient exploration, enabling the identification of top-performing candidates. 

\section{Conclusion}
In modern recommendation systems, approximately half of the performance improvements stem from refining the function that combines predictions during the value model stage. Despite this, existing auto-tuning systems fail to deliver viable solutions within the critical timeframe of 2-3 days. In this paper, we introduce an end-to-end auto-tuning system that efficiently and effectively optimizes multiple generic objectives and constraints by leveraging hourly system feedback. Our approach incorporates a novel hourly signal to address the challenges of noisy, non-i.i.d. data. We propose an efficient zeroth-order constrained optimization method that combines a Gaussian Process (GP) estimator with Thompson sampling (TS) to handle arbitrary problem forms. Additionally, our adoption of asynchronous parallel exploration significantly reduces the time required to find optimal solutions. 
This highlights the robustness of our approach and its potential for broader application in tuning tasks across various domains. 

\section*{Acknowledgements}
We would like to thank Chuanqi Wei, Yuhao Du, Yuting Zhang, Renjie Jiang, Yijia Liu, Bo Feng, Cheng Liu, Zellux Wang, Ashish Gandhe, Zhe Wang, Jieli Shen, Everest Chen, Syed Fahad Khalid, Lizhu Zhang, Dayong Wang, Pei Yin, Chao Yang, Hui Zhang, Shilin Ding and others for supporting this project.

\bibliographystyle{acm}
\balance
\bibliography{references}

\begin{thebibliography}{10}

\bibitem{emukit2018}
{\sc authors, T.~E.}
\newblock Emukit: Emulation and uncertainty quantification for decision making.
\newblock \url{https://github.com/amzn/emukit}, 2018.

\bibitem{gpyopt2016}
{\sc authors, T.~G.}
\newblock Gpyopt: A bayesian optimization framework in python.
\newblock \url{http://github.com/SheffieldML/GPyOpt}, 2016.

\bibitem{balandat2020botorch}
{\sc Balandat, M., Karrer, B., Jiang, D.~R., Daulton, S., Letham, B., Wilson, A.~G., and Bakshy, E.}
\newblock Botorch: A framework for efficient monte-carlo bayesian optimization.
\newblock In {\em Proceedings of Advances in Neural Information Processing Systems\/} (2020).

\bibitem{covington2016deep}
{\sc Covington, P., Adams, J., and Sargin, E.}
\newblock Deep neural networks for youtube recommendations.
\newblock In {\em Proceedings of the 10th ACM Conference on Recommender Systems\/} (2016).

\bibitem{dewancker2016bayesian}
{\sc Dewancker, I., McCourt, M., and Clark, S.}
\newblock Bayesian optimization for machine learning: A practical guidebook.
\newblock {\em arXiv preprint arXiv:1612.04858\/} (2016).

\bibitem{frazier2018bayesian}
{\sc Frazier, P.~I.}
\newblock Bayesian optimization.
\newblock In {\em Recent Advances in Optimization and Modeling of Contemporary Problems}. INFORMS, 2018, pp.~255--278.

\bibitem{galuzzi2020hyperparameter}
{\sc Galuzzi, B.~G., Giordani, I., Candelieri, A., Perego, R., and Archetti, F.}
\newblock Hyperparameter optimization for recommender systems through bayesian optimization.
\newblock {\em Computational Management Science 17\/} (2020), 495--515.

\bibitem{gao2022kuairand}
{\sc Gao, C., Li, S., Zhang, Y., Chen, J., Li, B., Lei, W., Jiang, P., and He, X.}
\newblock Kuairand: An unbiased sequential recommendation dataset with randomly exposed videos.
\newblock In {\em Proceedings of the 31st ACM International Conference on Information and Knowledge Management\/} (2022).

\bibitem{gomez2015netflix}
{\sc Gomez-Uribe, C.~A., and Hunt, N.}
\newblock The netflix recommender system: Algorithms, business value, and innovation.
\newblock {\em ACM Transactions on Management Information Systems (TMIS) 6}, 4 (2015), 1--19.

\bibitem{groh2012social}
{\sc Groh, G., Birnkammerer, S., and K{\"o}llhofer, V.}
\newblock Social recommender systems.
\newblock {\em Recommender Systems for the Social Web 32\/} (2012), 1.

\bibitem{tim_head_2018_1207017}
{\sc Head, T., MechCoder, Louppe, G., Shcherbatyi, I., fcharras, Vinícius, Z., cmmalone, Schröder, C., nel215, Campos, N., Young, T., Cereda, S., Fan, T., rene rex, Shi, K.~K., Schwabedal, J., carlosdanielcsantos, Hvass-Labs, Pak, M., SoManyUsernamesTaken, Callaway, F., Estève, L., Besson, L., Cherti, M., Pfannschmidt, K., Linzberger, F., Cauet, C., Gut, A., Mueller, A., and Fabisch, A.}
\newblock scikit-optimize/scikit-optimize: v0.5.2, Mar. 2018.

\bibitem{joglekar2020neural}
{\sc Joglekar, M.~R., Li, C., Chen, M., Xu, T., Wang, X., Adams, J.~K., Khaitan, P., Liu, J., and Le, Q.~V.}
\newblock Neural input search for large scale recommendation models.
\newblock In {\em Proceedings of the 26th ACM SIGKDD International Conference on Knowledge Discovery \& Data Mining\/} (2020).

\bibitem{kandasamy2020tuning}
{\sc Kandasamy, K., Vysyaraju, K.~R., Neiswanger, W., Paria, B., Collins, C.~R., Schneider, J., Poczos, B., and Xing, E.~P.}
\newblock Tuning hyperparameters without grad students: Scalable and robust bayesian optimisation with dragonfly.
\newblock {\em Journal of Machine Learning Research 21}, 81 (2020), 1--27.

\bibitem{klein2017robo}
{\sc Klein, A., Falkner, S., Mansur, N., and Hutter, F.}
\newblock Robo: A flexible and robust bayesian optimization framework in python.
\newblock In {\em NIPS 2017 Bayesian Optimization Workshop\/} (2017).

\bibitem{matuszyk2016comparative}
{\sc Matuszyk, P., Castillo, R.~T., Kottke, D., and Spiliopoulou, M.}
\newblock A comparative study on hyperparameter optimization for recommender systems.
\newblock In {\em Workshop on Recommender Systems and Big Data Analytics\/} (2016).

\bibitem{montanari2022impact}
{\sc Montanari, M., Bernardis, C., and Cremonesi, P.}
\newblock On the impact of data sampling on hyper-parameter optimisation of recommendation algorithms.
\newblock In {\em Proceedings of the 37th ACM/SIGAPP Symposium on Applied Computing\/} (2022).

\bibitem{moscati2023multiobjective}
{\sc Moscati, M., Deldjoo, Y., Carparelli, G.~D., and Schedl, M.}
\newblock Multiobjective hyperparameter optimization of recommender systems.
\newblock In {\em Perspectives@ RecSys\/} (2023).

\bibitem{shahriari2015taking}
{\sc Shahriari, B., Swersky, K., Wang, Z., Adams, R.~P., and De~Freitas, N.}
\newblock Taking the human out of the loop: A review of bayesian optimization.
\newblock {\em Proceedings of the IEEE 104}, 1 (2015), 148--175.

\bibitem{snoek2012practical}
{\sc Snoek, J., Larochelle, H., and Adams, R.~P.}
\newblock Practical bayesian optimization of machine learning algorithms.
\newblock In {\em Proceedings of Advances in Neural Information Processing Systems\/} (2012).

\bibitem{vanchinathan2014explore}
{\sc Vanchinathan, H.~P., Nikolic, I., De~Bona, F., and Krause, A.}
\newblock Explore-exploit in top-n recommender systems via gaussian processes.
\newblock In {\em Proceedings of the 8th ACM Conference on Recommender Systems\/} (2014).

\bibitem{wu2023hyperparameter}
{\sc Wu, D., Sun, B., and Shang, M.}
\newblock Hyperparameter learning for deep learning-based recommender systems.
\newblock {\em IEEE Transactions on Services Computing 16}, 4 (2023), 2699--2712.

\bibitem{wu2016parallel}
{\sc Wu, J., and Frazier, P.}
\newblock The parallel knowledge gradient method for batch bayesian optimization.
\newblock {\em Proceedings of Advances in Neural Information Processing Systems 29\/} (2016).

\bibitem{zheng2023automl}
{\sc Zheng, R., Qu, L., Cui, B., Shi, Y., and Yin, H.}
\newblock Automl for deep recommender systems: A survey.
\newblock {\em ACM Transactions on Information Systems 41}, 4 (2023), 1--38.

\end{thebibliography}
\allowdisplaybreaks[4]

\appendix
\balance
\section{Experiment Details}\label{app:exp_detail} 
\begin{figure}[H]
    \centering
    \hspace*{\fill}
    \subfigure[$\delta_1(\theta)$]{\includegraphics[width=0.23\textwidth]{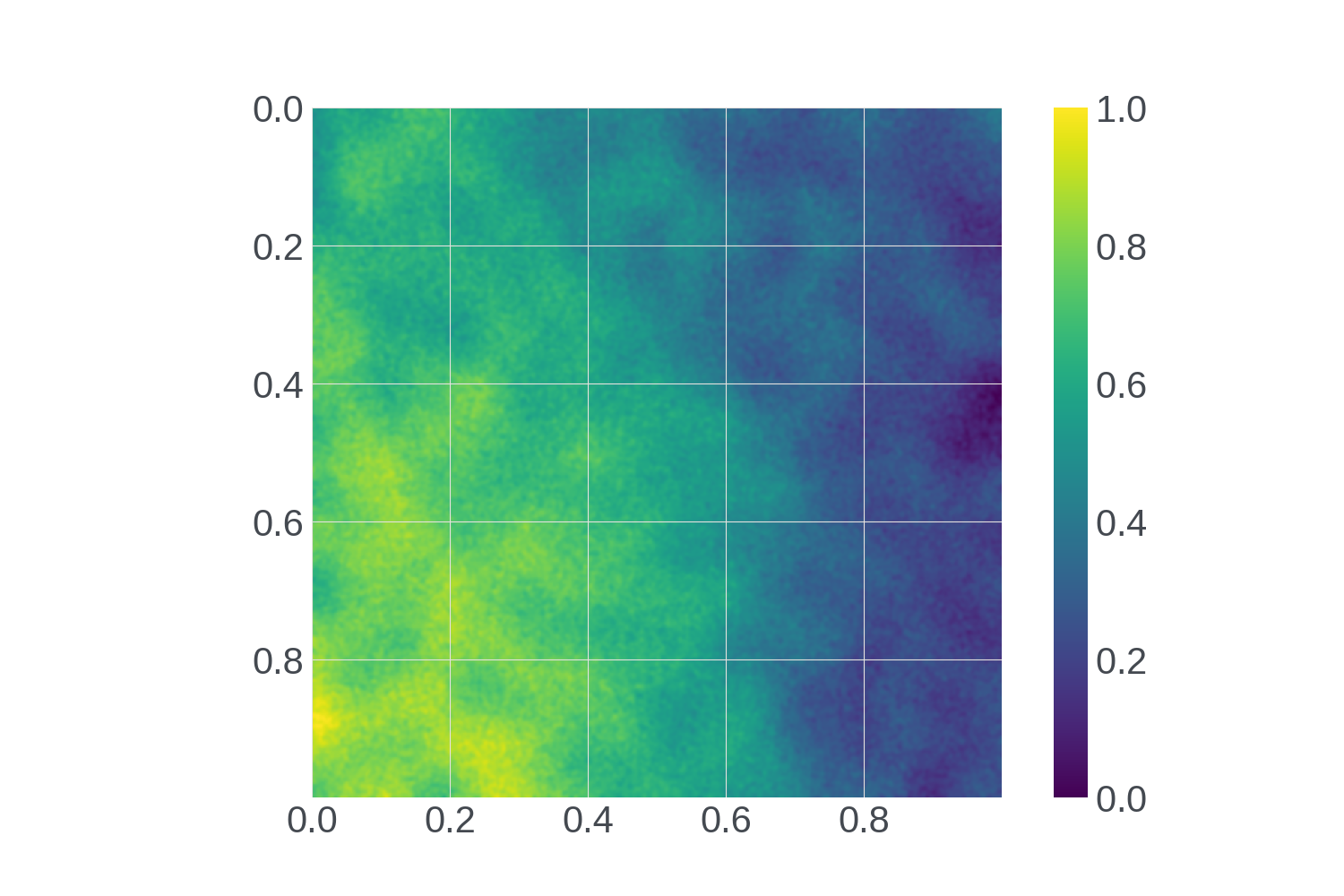}\label{fig:theta_1}}
    \hfill
    \subfigure[$\delta_2(\theta)$]{\includegraphics[width=0.23\textwidth]{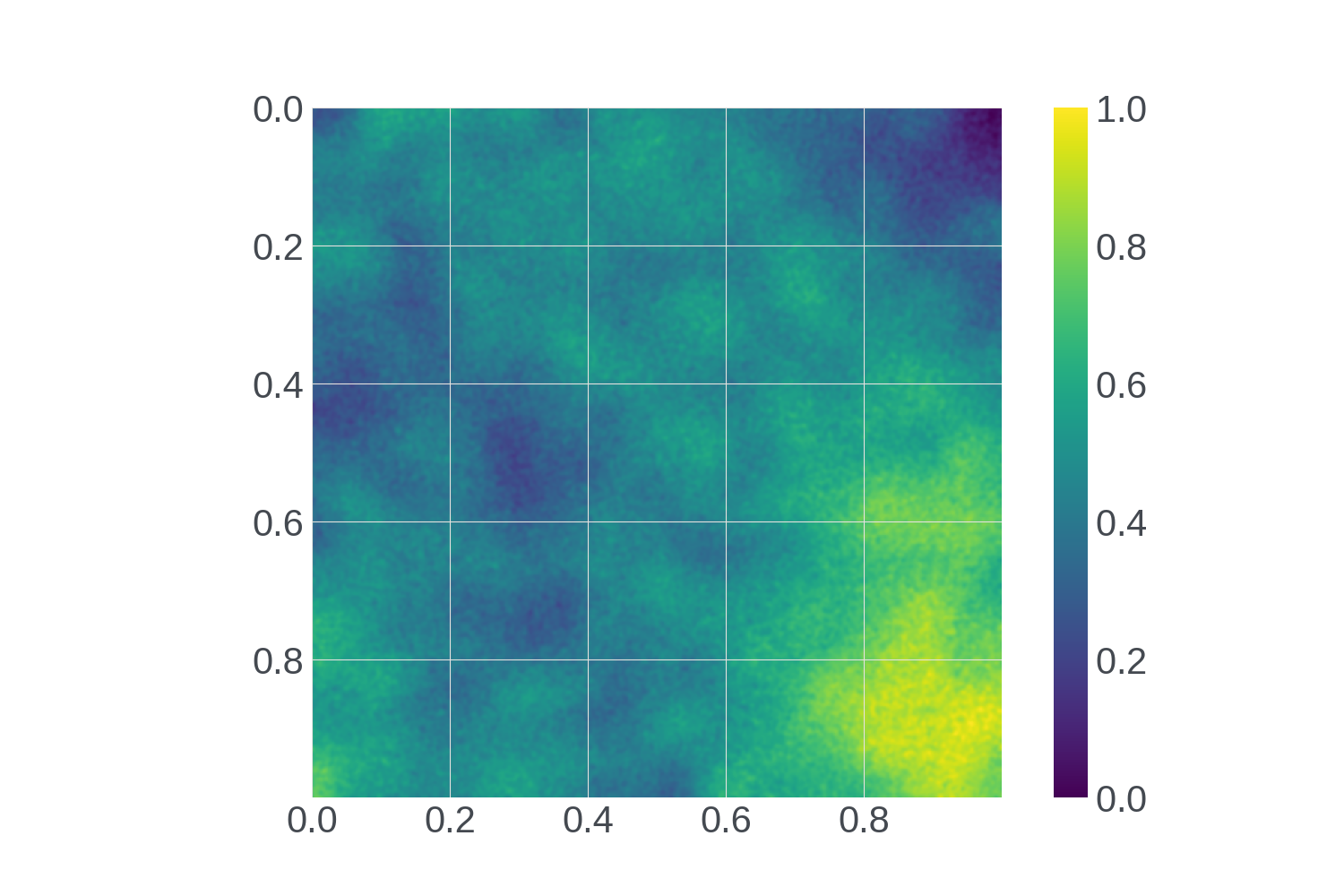}\label{fig:theta_2}}
    \hspace*{\fill}
    \caption{Randomly generated $\delta_1(\theta)$ and $\delta_2(\theta)$.}
    \label{fig:theta}
\end{figure}

\begin{figure}[H]
    \centering
    \includegraphics[width=0.24\textwidth]{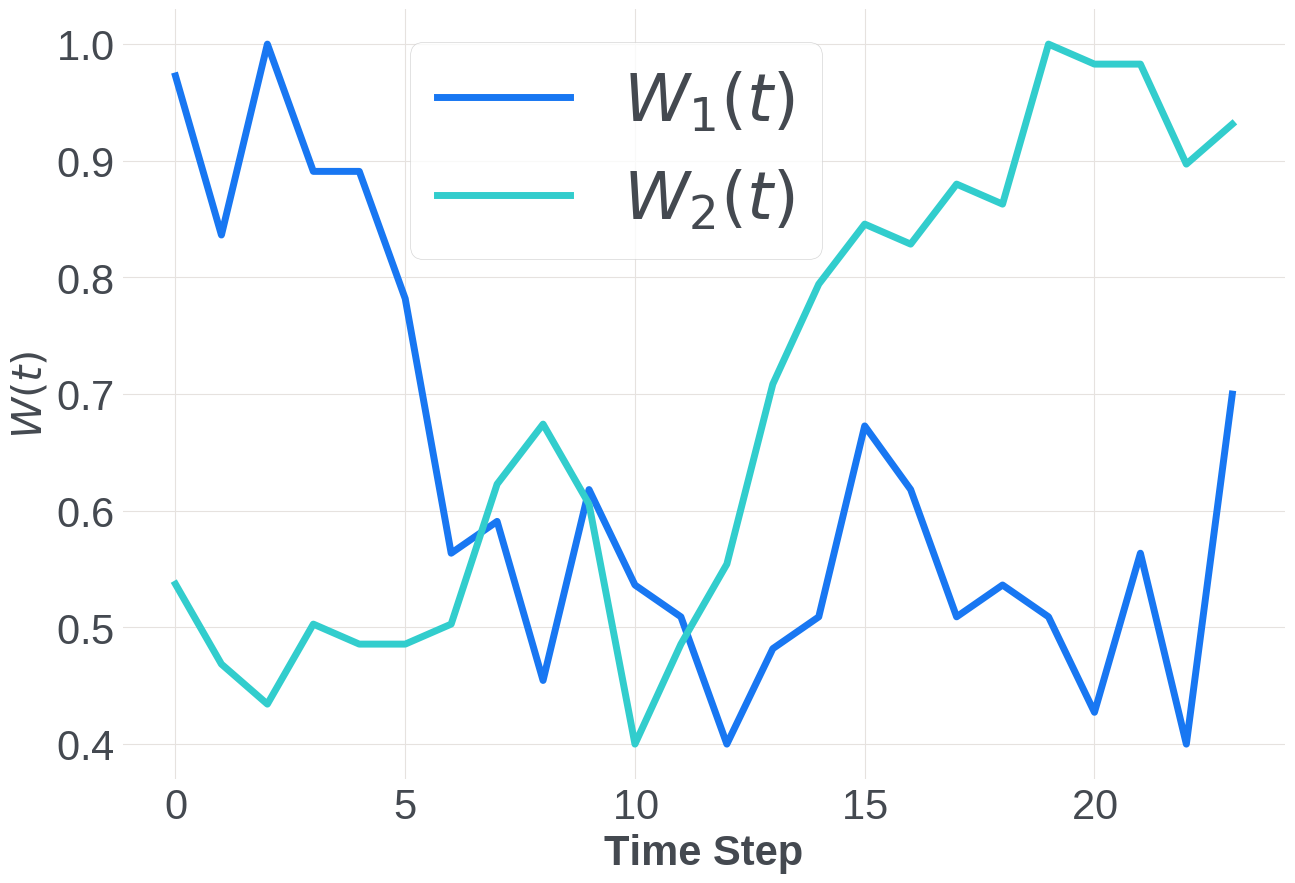}
    \caption{Simulated hourly metrics $W_1(t)$ and $W_2(t)$.}
    \label{fig:timefunction}
\end{figure}

In this section, we provide the experiment details on the synthetic environment in Section~\ref{sec:exp-synthetic}. We consider the following constrained optimization problem 
\begin{align*}
    \max_{\vtheta\in [0, 1]^2} \;& \Big\{f(\vtheta) := \vw_1 \mathbbm{E}\left[X_1(\vtheta; t)\right] + \vw_2 \mathbbm{E}\left[X_2(\vtheta; t)\right]\Big\} \\
    s.t. \;& g(\vtheta) := \vw_3\mathbbm{E}\left[X_1(\vtheta; t)\right] + \vw_4\mathbbm{E}\left[X_2(\vtheta; t)\right] \ge c,
\end{align*}
where $\vw = (0.296, 1.165, 0.149, 0.703)^\top $ and $c = 0.6036$ are randomly generated constants.
Meanwhile, $X_1(\vtheta; t)$ and $X_2(\vtheta; t)$ are two hourly metrics following the Gaussian distributions $\gN(\mu_1(\vtheta; t), \sigma^2)$ and $\gN(\mu_2(\vtheta; t), \sigma^2)$, respectively, where $\sigma = 0.6$ and
\begin{align*}
    \mu_1(\vtheta; t) \coloneqq \left(1 + 0.1 \delta_1(\vtheta)\right)W_1(t), \; \mu_2(\vtheta; t) \coloneqq \left(1 + 0.1 \delta_2(\vtheta)\right)W_2(t). 
\end{align*}
We randomly generate functions $\delta_1(\vtheta), \delta_2(\vtheta): [0, 1]^2 \rightarrow [0, 1]$ as in Figures~\ref{fig:theta_1}~and~\ref{fig:theta_2}, respectively. We choose $W_1(t)$ and $W_2(t)$ to be randomly generated periodic functions over $24$ time steps, which exhibit a strong trade-off with each other, motivated by true system metrics such as view counts and watch time, as illustrated in Figure~\ref{fig:timefunction}. For each time step, we draw $X_1(\vtheta; t)$ and $X_2(\vtheta; t)$ 50 times to simulate system metric collection. Moreover, we run the experiment over $50$ randomly chosen random seeds: $\{42,40,22,35,0,1,130,3,131,5,4,135,145,146,148,149,$ $61,151,21,28,$
$156,29,33,163,165,41,171,172,43,46,180,52,182,82,$
$183,185,187, $
$150,189,193,66,197,83,84,85,98,99,110,111,126\}$. We set the number of users available to be $10^6$, $T=30$, the number of proposed new candidates to be $600$, and the number of testing candidates to be $1000$. For the base policy, it is randomly picked as $\theta_{base} = (0.011, 0.985)^\top$.

\end{document}